# بررسی مقاومت الگوریتم مجموعه عضوی کمترین میانگین مربعات نرمال شده

رجب شعبانی
دانشگاه پیام نور

**Abstract**

In this letter, we analyze, two properties, the local and the global robustness of the set-membership normalized least mean square (SM-NLMS) algorithm. We will show that the SM-NLMS algorithm has $l_2$-stability. Indeed, the SM-NLMS algorithm never diverges; no matter how the parameters of the SM-NLMS algorithm has been selected. Ultimately, the numerical simulations corroborate the validity of our analysis.

چکیده

در این مقاله، دو ویژگی برای مقاومت الگوریتم مجموعه عضوی کمترین میانگین مربعات نرمال شده (SM-NLMS) ارائه می دهیم. ویژگی اول یک کران موضعی برای میزان خطا/اختلاف ضرایب در هر تکرار ارائه می دهد، در حالی که ویژگی دوم با این خاصیت را برای هر تکرار گسترش می دهد. در نتیجه این مقاله، برای اولین بار نشان می دهد که الگوریتم SM-NLMS فارغ از چگونگی انتخاب پارامترها از جهت پایداری با نرم $l_2$ مقاوم است.

## ۱ مقدمه

پردازش سیگنال تطبیقی یا به زبان ساده فیلترهای تطبیقی با ظهور الگوریتم کمترین میانگین مربعات ( LMS) به وجود آمدند [۱]. از آم زمان به بعد فیلترهای تطبیقی در بسیاری مسائل به

کار گرفته شدند، از جمله می توان شناسایی سیستمها، تقویت سیگنالها و پیش بینی سیگنالها را نام برد[۲، ۳] .

برای بهبود الگوریتم LMS، الگوریتم LMS نرمال شده (NLMS) در سال ۱۹۶۷ ارائه گردید [۴، ۵]. در واقع NLMS مشکل نویز را در گرادیانت حل می کند، و انتخاب پارامتر همگرایی را در مقایسه با الگوریتم LMS بدون هزینه محاسباتی بالا بسیار آسان می کند. به همین دلیل استفاده از الگوریتم NLMS بسیار مرسوم تر است [۶، ۷، ۸، ۹].

در سال ۱۹۹۸، نسخه داده-انتخابی الگوریتم NLMS، با نام الگوریتم مجموعه عضوی کمترین میانگین مربعات نرمال شده (SM-NLMS) ارائه گردید [۱۰]. الگوریتم SM-NLMS با حفظ مزیت های الگوریتم NLMS، به کمک خاصیت داده-انتخابی می تواند قبل از استفاده از داده های جدید آنها را ارزیابی کند و در صورت مفید بودن آنها را در فرایند یادگیری مورد استفاده قرار دهد. بدین وسیله الگوریتم SM-NLMS در مقایسه با NLMS دارای دقت بیشتر و مقاومت بیشتر در برابر نویز است و همچنین هزینه محاسباتی کمتری را برای بروزرسانی ضرایب صرف می کند زیرا تنها در صورتی داده های جدید را در فرایند یادگیری به کار می گیرد که در آنها اطلاعات جدیدی موجود باشد [۱۱، ۱۲، ۱۳].

با وجود تمام مزیت هایی که برای الگوریتم SM-NLMS وجود دارد، اما این الگوریتم هنوزه در صنعت به اندازه کافی مورد استفاده قرار نمی گیرد، دلیل آن می تواند کمبود بررسی ویژگی های تئوری این الگوریتم باشد [۱۴]. در این مقاله، پایداری الگوریتم SM-NLMS را از نظر نرم $l_2$ مورد بررسی قرار می دهیم. در بخش ۲ کمی در مورد الگوریتم های مجموعه عضوی توضیح می دهیم و در بخش ۳ الگوریتم SM-NLMS را مرور می کنیم. سپس در بخش ۴ دو خاصیت کلی را برای میزان پایداری این الگوریتم مورد بحث قرار می دهیم. در بخش ۵ شبیه سازی و آزمایش ها را ارائه می کنیم و در انتها نتیجه گیری در بخش ۶ جمع بندی می شود.

نمادها: برای تکرار $k$، ضرایب فیلتر تطبیقی و بردار ورودی به ترتیب با $\mathbf{w}(k)$ و $\mathbf{x}(k)$ در $\mathrm{R}^{N+1}$ نشان داده می شوند. سیگنال های مرجع، خروجی و خطا به ترتیب با $y(k), d(k)$ و $e(k)$ در $\mathrm{R}$ نمایش داده می شوند که $y(k) = \mathbf{x}^T(k)\mathbf{w}(k)$ و $e(k) = d(k) - y(k)$.

## ۲ فیلترهای مجموعه عضوی

هدف از فیلترهای مجموعه عضوی محاسبه ضرایب $\mathbf{w} \in \mathrm{R}^{N+1}$ به گونه ای می باشد که اندازه خطای خروجی توسط یک مقدار از قبل تعیین شده $\overline{\gamma} \in \mathrm{R}_+$ محدود شده باشد. مقدار $\overline{\gamma}$ بسته به کاربردهای مورد نظر می تواند متفاوت باشد. اگر مقدار $\overline{\gamma}$ به طور مناسب انتخاب شده باشد، مقادیر بسیار معتبری برای $\mathbf{w}$ وجود خواهد داشت. فرض کنید $S$ مجموعه تمامی مقادیر ممکن برای جفت داده سیگنال ورودی و سیگنال مرجع $(\mathbf{x}, d)$ و $\Theta$ مجموعه تمامی بردارهای $\mathbf{w}$ باشد که اندازه خطای خروجی آنها کمتر از $\overline{\gamma}$ است. مجموعه $\Theta$، مجموعه جواب های پذیرفتنی

نامیده می شود، و به شکل زیر تعریف می شود،

$$\Theta = \bigcap_{(\mathbf{x},\mathrm{d})\in\mathcal{S}} \{\mathbf{w} \in \mathrm{R}^{\mathrm{N}+1} : |\mathrm{d} - \mathbf{w}^{\mathrm{T}}\mathbf{x}| \leq \overline{\gamma}\}. \qquad (۱)$$

فرض کنید مجموعه محدوده $\mathcal{H}(\mathrm{k})$ شامل همه بردارهای $\mathbf{w}$ در لحظه تکرار $\mathrm{k}$ باشد که اندازه خطای خروجی آنها کمتر از $\overline{\gamma}$ است، یعنی

$$\mathcal{H}(\mathrm{k}) = \{\mathbf{w} \in= \mathrm{R}^{\mathrm{N}+1} : |\mathrm{d}(\mathrm{k}) - \mathbf{w}^{\mathrm{T}}\mathbf{x}(\mathrm{k})| \leq \overline{\gamma}\}. \qquad (۲)$$

مجموعه عضوی $\psi(\mathrm{k})$ به گونه زیر تعریف می شود

$$\psi(\mathrm{k}) = \bigcap_{\mathrm{i}=0}^{\mathrm{k}} \mathcal{H}(\mathrm{i}), \qquad (۳)$$

که شامل $\Theta$ می باشد، و اگر همه جفتهای موجود $\mathcal{S}$ را پردازش کند، دقیقا برابر $\Theta$ می شود.به خاطر سختی محاسبه $\psi(\mathrm{k})$ روش های محاسبه تکراری مورد استفاده قرار می گیرد [۱۰]. آسان ترین روش، مانند الگوریتم کمترین میانگین مربعات نرمال شده، استفاده از اطلاعات موجود در $\mathcal{H}(\mathrm{k})$ است [۱۰]، یا می توان چندین مجموعه محدوده آخر را مانند الگوریتم انعکاس آفین در نظر گرفت [۱۵].

# ۳ الگوریتم مجموعه عضوی کمترین میانگین مربعات نرمال شده (SM-NLMS)

الگوریتم SM-NLMS با معادله بازگشتی زیر ارائه می شود [۲]،

$$\mathbf{w}(\mathrm{k}+1) = \mathbf{w}(\mathrm{k}) + \frac{\mu(\mathrm{k})}{\mathbf{x}(\mathrm{k})^2 + \delta}\mathrm{e}(\mathrm{k})\mathbf{x}(\mathrm{k}), \qquad (۴)$$

که در آن

$$\mu(\mathrm{k}) = \begin{cases} 1 - \frac{\overline{\gamma}}{|\mathrm{e}(\mathrm{k})|} & \text{اگر}\ |\mathrm{e}(\mathrm{k})| > \overline{\gamma}, \\ 0 & \text{در غیر اینصورت}, \end{cases} \qquad (۵)$$

و $\overline{\gamma} \in \mathrm{R}_+$ یک کران بالای مثبت برای سیگنال خطای خروجی قابل قبول می باشد و معمولا به عنوان ضریبی از انحراف معیار سیگنال نویز انتخاب می شود [۲]. پارامتر یک عددمثبت بسیار کوچک است که برای جلوگیری از تقسیم بر صفر برگزیده شده است.

## ۴ مقاومت الگوریتم SM-NLMS

یک سناریوی شناسایی سیستم را در نظر بگیرید که در آن سیستم ناشناخته توسط $\mathbf{w}_o \in \mathrm{R}^{N+1}$ و سیگنال مرجع توسط $\mathrm{d(k)}$ نمایش داده می شودو به گونه زیر تعریف شده است

$$\mathrm{d(k)} = \mathbf{w}_o^T \mathbf{x}(\mathrm{k}) + \mathrm{n(k)}, \tag{۶}$$

که در آن $\mathrm{n(k)} \in \mathrm{R}$ سیگنال نویز را نشان می دهد. یکی از مشکلات اساسی در تحلیل مقاومت الگوریتم SM-NLMS وضعیت شرطی معادله (۵) می باشد. برای رفع این مشکل می توانیم

$$\overline{\mu}(\mathrm{k}) = 1 - \frac{\overline{\gamma}}{|\mathrm{e(k)}|} \tag{۷}$$

و تابع مشخصه $\mathrm{f} : \mathrm{R} \times \mathrm{R}_+ \rightarrow \{0, 1\}$ را به گونه زیر تعریف کنیم

$$\mathrm{f}(\mathrm{e(k)}, \overline{\gamma}) = \begin{cases} 1 & \text{اگر} |\mathrm{e(k)}| > \overline{\gamma}, \\ 0 & \text{در عیر اینصورت.} \end{cases} \tag{۸}$$

بدین ترتیب، معادله بازگشتی الگوریتم SM-NLMS به صورت زیر بازنویسی می شود

$$\mathbf{w}(\mathrm{k}+1) = \mathbf{w}(\mathrm{k}) + \frac{\overline{\mu}(\mathrm{k})}{\alpha(\mathrm{k})} \mathrm{e(k)} \mathbf{x}(\mathrm{k}) \mathrm{f}(\mathrm{e(k)}, \overline{\gamma}), \tag{۹}$$

که در آن

$$\alpha(\mathrm{k}) = \mathbf{x}(\mathrm{k})^2 + \delta. \tag{۱۰}$$

چون هدف ما مطالعه خاصیت مقاومت الگوریتم است، تعریف زیر از $\widetilde{\mathbf{w}}(\mathrm{k}) \in \mathrm{R}^{N+1}$ مفید خواهد بود

$$\widetilde{\mathbf{w}}(\mathrm{k}) = \mathbf{w}_o - \mathbf{w}(\mathrm{k}), \tag{۱۱}$$

که $\widetilde{\mathbf{w}}(\mathrm{k})$ نشاندهنده تفاوت بین $\mathbf{w}_o$ و بردار $\mathbf{w}(\mathrm{k})$ است که به دنبال تخمین آن هستیم. بنابراین سیگنال خطا به گونه زیر بازنویسی خواهد شد

$$\begin{aligned} \mathrm{e(k)} &= \mathrm{d(k)} - \mathbf{w}^T(\mathrm{k})\mathbf{x}(\mathrm{k}) = \mathbf{w}_o^T \mathbf{x}(\mathrm{k}) + \mathrm{n(k)} - \mathbf{w}^T(\mathrm{k})\mathbf{x}(\mathrm{k}) \\ &= \underbrace{\widetilde{\mathbf{w}}^T(\mathrm{k})\mathbf{x}(\mathrm{k})}_{=\widetilde{\mathrm{e}}(\mathrm{k})} + \mathrm{n(k)}, \end{aligned} \tag{۱۲}$$

که $\widetilde{\mathrm{e}}(\mathrm{k})$ خطای بدون نویز می باشد.

با جایگزینی (۱۱) در (۹) خواهیم داشت

$$\widetilde{\mathbf{w}}(k+1) = \widetilde{\mathbf{w}}(k) - \frac{\overline{\mu}(k)}{\alpha(k)} e(k)\mathbf{x}(k) f(e(k), \overline{\gamma}), \quad (۱۳)$$

که با تجزیه $e(k)$ از (۱۲) خواهیم داشت

$$\widetilde{\mathbf{w}}(k+1) = \widetilde{\mathbf{w}}(k) - \frac{\overline{\mu}(k)}{\alpha(k)} \widetilde{e}(k)\mathbf{x}(k) f(e(k), \overline{\gamma}) - \frac{\overline{\mu}(k)}{\alpha(k)} n(k)\mathbf{x}(k) f(e(k), \overline{\gamma}). \quad (۱۴)$$

بعد یک سری محاسبات ریاضی در معادله بالا، قضیه زیر را برای خاصیت مقاومت الگوریتم SM-NLMS به دست می آوریم.

قضیه ۱ (مقاومت موضعی الگوریتم SM-NLMS): برای الگوریتم SM-NLMS همواره داریم

$$\widetilde{\mathbf{w}}(k+1)^2 = \widetilde{\mathbf{w}}(k)^2, \text{ □□ } f(e(k), \overline{\gamma}) = 0 \quad (۱۵)$$

یا

$$\widetilde{\mathbf{w}}(k+1)^2 + \frac{\overline{\mu}(k)}{\alpha(k)} \widetilde{e}^2(k) < \widetilde{\mathbf{w}}(k)^2 + \frac{\overline{\mu}(k)}{\alpha(k)} n^2(k) \;, \quad (۱۶)$$

اگر $f(e(k), \overline{\gamma}) = 1$.

اثبات: ابتدا برای آسانی کار اندیس $k$ و پارامترهای تابع $f$ را حذف میکنیم. پس داریم

$$\widetilde{\mathbf{w}}(k+1) = \widetilde{\mathbf{w}} - \frac{\overline{\mu}}{\alpha} \widetilde{e}\mathbf{x}f - \frac{\overline{\mu}}{\alpha} n\mathbf{x}f. \quad (۱۷)$$

با بدست آوردن نرم دو طرف خواهیم داشت

$$\begin{aligned}
\widetilde{\mathbf{w}}(k+1)^2 =& \widetilde{\mathbf{w}}^T\widetilde{\mathbf{w}} - \frac{\overline{\mu}}{\alpha}\widetilde{e}\widetilde{\mathbf{w}}^T\mathbf{x}f - \frac{\overline{\mu}}{\alpha}n\widetilde{\mathbf{w}}^T\mathbf{x}f - \frac{\overline{\mu}}{\alpha}\widetilde{e}\mathbf{x}^T\widetilde{\mathbf{w}}f + \frac{\overline{\mu}^2}{\alpha^2}\widetilde{e}^2\mathbf{x}^T\mathbf{x}f^2 \\
&+ \frac{\overline{\mu}^2}{\alpha^2}\widetilde{e}n\mathbf{x}^T\mathbf{x}f^2 - \frac{\overline{\mu}}{\alpha}n\mathbf{x}^T\widetilde{\mathbf{w}}f + \frac{\overline{\mu}^2}{\alpha^2}n\widetilde{e}\mathbf{x}^T\mathbf{x}f^2 + \frac{\overline{\mu}^2}{\alpha^2}n^2\mathbf{x}^T\mathbf{x}f^2 \\
=& \widetilde{\mathbf{w}}^2 - \frac{\overline{\mu}}{\alpha}\widetilde{e}^2f - \frac{\overline{\mu}}{\alpha}n\widetilde{e}f - \frac{\overline{\mu}}{\alpha}\widetilde{e}^2f + \frac{\overline{\mu}^2}{\alpha^2}\widetilde{e}^2\mathbf{x}^2f^2 + \frac{\overline{\mu}^2}{\alpha^2}\widetilde{e}n\mathbf{x}^2f^2 \\
&- \frac{\overline{\mu}}{\alpha}n\widetilde{e}f + \frac{\overline{\mu}^2}{\alpha^2}n\widetilde{e}\mathbf{x}^2f^2 + \frac{\overline{\mu}^2}{\alpha^2}n^2\mathbf{x}^2f^2 \\
=& \widetilde{\mathbf{w}}^2 - 2\frac{\overline{\mu}}{\alpha}\widetilde{e}^2f - 2\frac{\overline{\mu}}{\alpha}n\widetilde{e}f + (\widetilde{e}+n)^2\frac{\overline{\mu}^2}{\alpha^2}\mathbf{x}^2f^2 \\
=& \widetilde{\mathbf{w}}^2 + (\widetilde{e}+n)^2\frac{\overline{\mu}^2}{\alpha^2}\mathbf{x}^2f^2 - 2\frac{\overline{\mu}}{\alpha}\widetilde{e}^2f - 2\frac{\overline{\mu}}{\alpha}n\widetilde{e}f - \frac{\overline{\mu}}{\alpha}n^2f + \frac{\overline{\mu}}{\alpha}n^2f \\
=& \widetilde{\mathbf{w}}^2 + (\widetilde{e}+n)^2\frac{\overline{\mu}^2}{\alpha^2}\mathbf{x}^2f^2 + \frac{\overline{\mu}}{\alpha}n^2f - (\widetilde{e}+n)^2\frac{\overline{\mu}}{\alpha}f - \frac{\overline{\mu}}{\alpha}\widetilde{e}^2f, \qquad (۱۸)
\end{aligned}$$

که تساوی دوم به دلیل $\widetilde{e} = \widetilde{\mathbf{w}}^T\mathbf{x} = \mathbf{x}^T\widetilde{\mathbf{w}}$ به دست آمده است. با مرتب کردن (۱۸) خواهیم داشت

$$\widetilde{\mathbf{w}}(k+1)^2 + \frac{\overline{\mu}f}{\alpha}\widetilde{e}^2 = \widetilde{\mathbf{w}}^2 + \frac{\overline{\mu}f}{\alpha}n^2 + c_1c_2, \qquad (۱۹)$$

که در آن

$$c_1 = \frac{\overline{\mu}f}{\alpha}(\widetilde{e}+n)^2, \qquad (۲۰)$$

$$c_2 = \frac{\overline{\mu}f}{\alpha}\mathbf{x}^2 - 1. \qquad (۲۱)$$

از (۱۹) زمانی که $f = 0$ داریم

$$\widetilde{\mathbf{w}}(k+1)^2 = \widetilde{\mathbf{w}}(k)^2 \qquad (۲۲)$$

همان طور که انتظار داریم زیرا $f = 0$ یعنی هیچ بروزرسانی صورت نگرفته است. اما وقتی $f = 1$ داریم $1 > \overline{\mu} > 0$ و $(\widetilde{e}+n)^2 = e^2 > \overline{\gamma}^2 > 0$.به علاوه مشاهده می کنیم که به خاطر معادله (۱۰) و $\delta > 0$ داریم $0 \leq \mathbf{x}^2/\alpha < 1$. با ترکیب این نامساوی ها خواهیم داشت $c_2 < 0$ و $c_1 > 0$. بنابراین زمانی که $f = 1$ داریم $c_1c_2 < 0$ که نامعادله زیر را نتیجه می دهد

$$\widetilde{\mathbf{w}}(k+1)^2 + \frac{\overline{\mu}}{\alpha}\widetilde{e}^2 < \widetilde{\mathbf{w}}^2 + \frac{\overline{\mu}}{\alpha}n^2. \qquad (۲۳)$$

اگر اندیس $\mathrm{k}$ را برگردانیم برای $f(e(k), \overline{\gamma}) = 1$ خواهیم داشت

$$\widetilde{\mathbf{w}}(k+1)^2 + \frac{\overline{\mu}(k)}{\alpha(k)}\widetilde{e}^2(k) < \widetilde{\mathbf{w}}(k)^2 + \frac{\overline{\mu}(k)}{\alpha(k)}n^2(k). \qquad \square \tag{۲۴}$$

این قضیه یک کران موضعی برای انحراف ضرایب از یک تکرار به تکرار بعدی ارایه می دهد. در واقع (۱۵) بیان می کند زمانی که هیچ بروزرسانی صورت نگرفته است انحراف ضرایب هیچ تغییری نمی کند. اما زمانی که بروزرسانی صورت می گیرد (۱۶) یک کران برای $\widetilde{\mathbf{w}}(k+1)^2$ بر اساس $\widetilde{\mathbf{w}}(k)^2$، $\widetilde{e}^2(k)$ و $n^2(k)$ ارایه می دهد. علاوه بر این تبصره زیر از قضیه بالا برای مقاومت سراسری ارایه داده می شود.

تبصره (مقاومت سراسری الگوریتم SM-NLMS): فرض کنید که الگوریتم SM-NLMS از لحظه ۰ (مقداردهی اولیه) تا لحظه $\mathrm{K}$ اجرا می شود. آنگاه خواهیم داشت

$$\frac{\widetilde{\mathbf{w}}(K)^2 + \sum_{k\in\mathcal{K}_{up}} \frac{\overline{\mu}(k)}{\alpha(k)}\widetilde{e}^2(k)}{\widetilde{\mathbf{w}}(0)^2 + \sum_{k\in\mathcal{K}_{up}} \frac{\overline{\mu}(k)}{\alpha(k)}n^2(k)} < 1, \tag{۲۵}$$

که $\mathcal{K}_{up}$ مجموعه تکرارهایی است که بروزرسانی $\mathbf{w}(k)$ صورت گرفته است.

اثبات: مجموعه همه تکرارهای تحت آتالیز را با $\mathcal{K} = \{0, 1, 2, \ldots, K-1\}$ نمایش دهید. $\mathcal{K}_{up}$ زیرمجموعه $\mathcal{K}$ شامل تکرارهایی است که در آنها بروزرسانی صورت گرفته است و $\mathcal{K}_{up}^c = \mathcal{K} \setminus \mathcal{K}_{up}$ مجموعه تکرارهایی است که در آنها بروزرسانی صورت نگرفته است. از قضیه ۱ هنگامی که $\mathbf{w}(k)$ بروزرسانی شده است (۱۶) برقرار است. با جمع کردن این نامساوی برای همه $k \in \mathcal{K}_{up}$ خواهیم داشت

$$\sum_{k\in\mathcal{K}_{up}} \left(\widetilde{\mathbf{w}}(k+1)^2 + \frac{\overline{\mu}(k)}{\alpha(k)}\widetilde{e}^2(k)\right) < \sum_{k\in\mathcal{K}_{up}} \left(\widetilde{\mathbf{w}}(k)^2 + \frac{\overline{\mu}(k)}{\alpha(k)}n^2(k)\right). \tag{۲۶}$$

به طور مشابه برای همه $k \in \mathcal{K}_{up}^c$ با استفاده از (۱۵) داریم

$$\sum_{k\in\mathcal{K}_{up}^c} \mathbf{w}(k+1)^2 = \sum_{k\in\mathcal{K}_{up}^c} \mathbf{w}(k)^2. \tag{۲۷}$$

با ترکیب کردن (۲۶) و (۲۷) داریم

$$\sum_{k\in\mathcal{K}} \widetilde{\mathbf{w}}(k+1)^2 + \sum_{k\in\mathcal{K}_{up}} \frac{\overline{\mu}(k)}{\alpha(k)}\widetilde{e}^2(k) < \sum_{k\in\mathcal{K}} \widetilde{\mathbf{w}}(k)^2 + \sum_{k\in\mathcal{K}_{up}} \frac{\overline{\mu}(k)}{\alpha(k)}n^2(k). \tag{۲۸}$$

اما مقادیر زیادی از $\widetilde{\mathrm{w}}(\mathrm{k})^2$ از دو طرف (۲۸) حذف می شوند، و معادله به حالت زیر ساده می شود

$$\widetilde{\mathrm{w}}(\mathrm{K})^2+\sum_{\mathrm{k}\in\mathcal{K}_{\mathrm{up}}}\frac{\overline{\mu}(\mathrm{k})}{\alpha(\mathrm{k})}\widetilde{\mathrm{e}}^2(\mathrm{k})<\widetilde{\mathrm{w}}(0)^2+\sum_{\mathrm{k}\in\mathcal{K}_{\mathrm{up}}}\frac{\overline{\mu}(\mathrm{k})}{\alpha(\mathrm{k})}\mathrm{n}^2(\mathrm{k}) \tag{۲۹}$$

با فرض غیر صفر بودت مخرج خواهیم داشت

$$\frac{\widetilde{\mathrm{w}}(\mathrm{K})^2+\sum_{\mathrm{k}\in\mathcal{K}_{\mathrm{up}}}\frac{\overline{\mu}(\mathrm{k})}{\alpha(\mathrm{k})}\widetilde{\mathrm{e}}^2(\mathrm{k})}{\widetilde{\mathrm{w}}(0)^2+\sum_{\mathrm{k}\in\mathcal{K}_{\mathrm{up}}}\frac{\overline{\mu}(\mathrm{k})}{\alpha(\mathrm{k})}\mathrm{n}^2(\mathrm{k})}<1. \qquad \square \tag{۳۰}$$

این تبصره نشان می دهد که مقاومت $\mathrm{l}_2$ الگوریتم SM-NLMS از مقادیر نااطمینانی $\{\widetilde{\mathrm{w}}(0),\{\mathrm{n}(\mathrm{k})\}_{0\leq\mathrm{k}\leq\mathrm{K}}\}$ به خطاهای $\{\widetilde{\mathrm{w}}(\mathrm{K}),\{\widetilde{\mathrm{e}}(\mathrm{k})\}_{0\leq\mathrm{k}\leq\mathrm{K}}\}$ همواره تضمین شده است. برخلاف الگوریتم NLMS که برای مقاومت $\mathrm{l}_2$ مقدار گام باید به دقت انتخاب شود، الگوریتم SM-NLMS این مزیت تضمین شده است.

# ۵ شبیه سازی

در این بخش الگوریتم SM-NLMS برای شناسایی سیستم به کار برده می شود. سیستم ناشناخته ۱۰ مولفه دارد که از توزیع نرمال استاندارد گرفته شده است. سیگنال ورودی یک سیگنال BPSK با واریانس ۱ است. نسبت نویز به سیگنال برابر ۲۰ دسیبل است. همچنین $\sigma_{\mathrm{n}}^2=0.01$ ، $\delta=10^{-12}$ ، $\mathbf{w}(0)=[0\ \cdots\ 0]^{\mathrm{T}}$ و $\overline{\gamma}=\sqrt{\tau\sigma_{\mathrm{n}}^2}$ که در آن $\tau\in\{2,5\}$. علاوه بر این سمت چپ و راست (۱۶) را به ترتیب با $\mathrm{g}_1(\mathrm{k})$ و $\mathrm{g}_2(\mathrm{k})$ نشان می دهیم.

شکل ۱ نتایج $\mathrm{g}_1(\mathrm{k})$ و $\mathrm{g}_2(\mathrm{k})$ را نشان می دهد. در این شکل برای هر دو مقدار $\overline{\gamma}$ شاهد هستیم که $\mathrm{g}_1(\mathrm{k})$ اکیدا زیر $\mathrm{g}_2(\mathrm{k})$ قرار دارد یا روی همدیگر قرار گرفته اند، یعنی $\mathrm{g}_1(\mathrm{k})\leq\mathrm{g}_2(\mathrm{k})$. در واقع زمانی که $\overline{\mu}(\mathrm{k})=0$، الگوریتم SM-NLMS هیچ بروزرسانی انجام نمی دهد و داریم $\mathrm{g}_1(\mathrm{k})=\mathrm{g}_2(\mathrm{k})$، در غیر این صورت $\mathrm{g}_1(\mathrm{k})<\mathrm{g}_2(\mathrm{k})$، همان گونه که در قضیه ۱ ادعا شده است.

# ۶ نتایج

در این مقاله مقاومت الگوریتم SM-NLMS بررسی شد. در ابتدا مقاومت موضعی مورد بررسی قرار گرفت و سپس به کمک آن مقاومت سراسری بررسی شد. در نتیجه در الگوریتم SM-NLMS فارغ از نحوه انتخاب پارامترهای الگوریتم، همواره انرژی خطا از انرژی نا اطمینانی ها

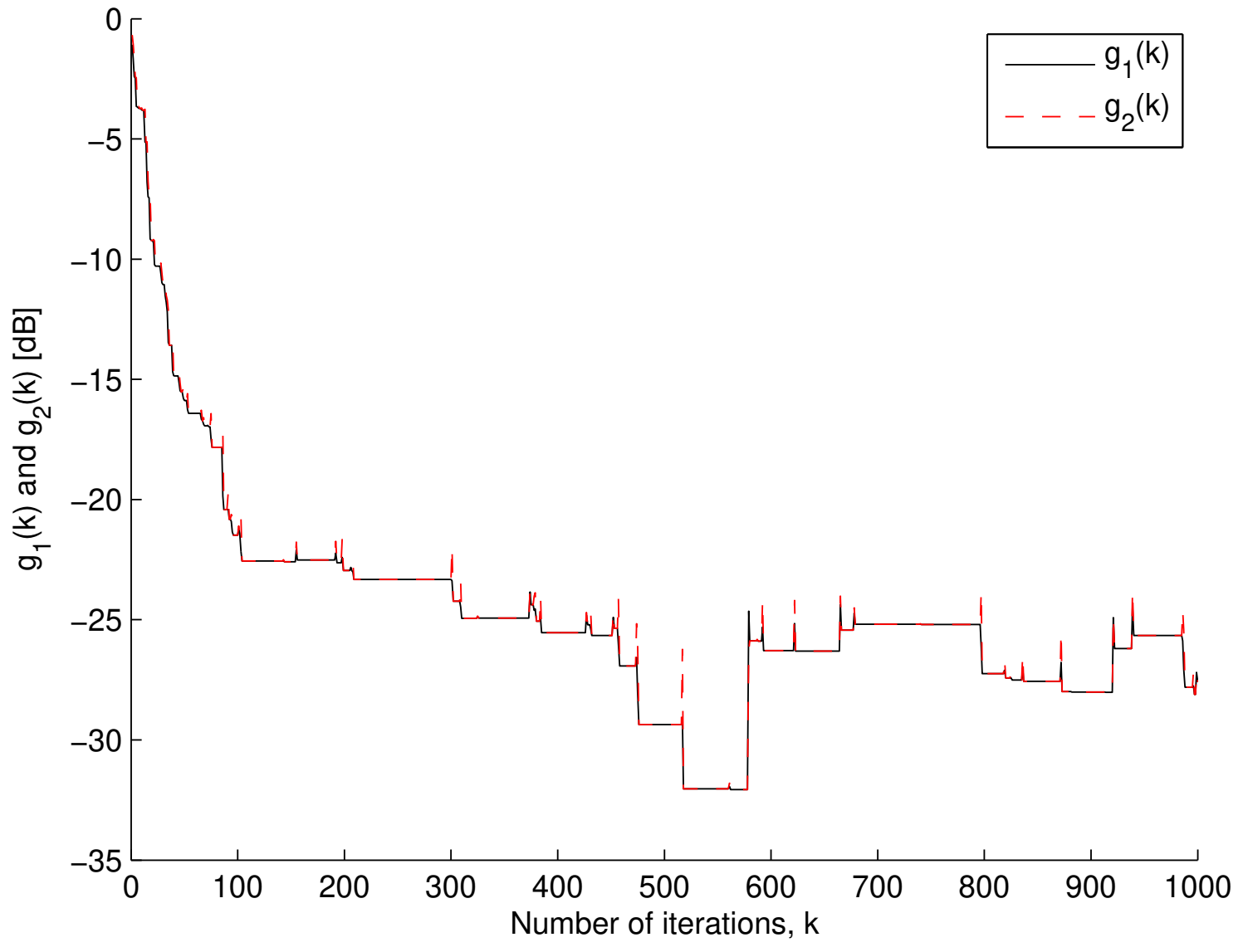

(آ) $\overline{\gamma} = \sqrt{5\sigma_n^2}$

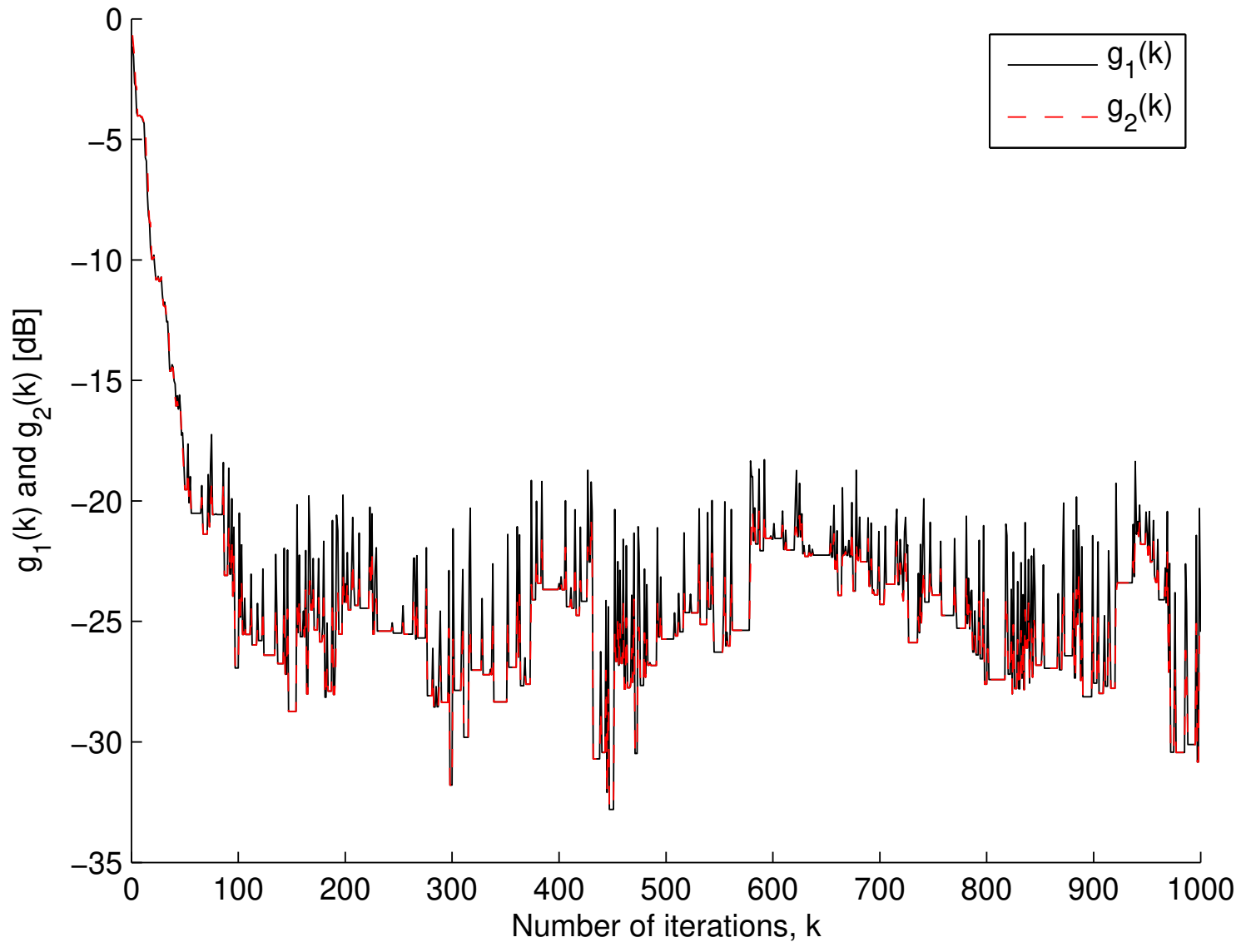

(ب) $\overline{\gamma} = \sqrt{2\sigma_n^2}$

شکل ۱: مقادیر $g_1(k)$ و $g_2(k)$.

کمتر است. پس علاوه بر مزیتهای ذکر شده برای الگوریتم SM-NLMS نسبت به الگوریتم NLMS، این الگوریتم از نظر نرم $l_2$ مقاوم است.

# مراجع